# The role of the Wigner distribution function in iterative ptychography

T B Edo[1]

[1] Department of Physics and Astronomy, University of Sheffield, S3 7RH, United Kingdom

E-mail: tega.edo@sheffield.ac.uk

Ptychography employs a set of diffraction patterns that capture redundant information about an illuminated specimen as a localized beam is moved over the specimen. The robustness of this method comes from the redundancy of the dataset that in turn depends on the amount of oversampling and the form of the illumination. Although the role of oversampling in ptychography is fairly well understood, the same cannot be said of the illumination structure. This paper provides a vector space model of ptychography that accounts for the illumination structure in a way that highlights the role of the Wigner distribution function in iterative ptychography.

## 1.     Introduction

Iterative phase retrieval (IPR) algorithms provide a very efficient way to recover the specimen function from a set of diffraction patterns in ptychography. The on-going developments behind the execution of these algorithms are pushing the boundaries of the method by extracting more information from the diffraction patterns than was earlier thought possible when iterative ptychography was first proposed [1]. Amongst these are: simultaneous recovery of the illumination (or probe) alongside the specimen [2, 3, 4], super resolved specimen recovery [5, 6], probe position correction [2, 7, 8, 9], three-dimensional specimen recovery [10, 11, 23], ptychography using under sampled diffraction patterns [12, 13, 14] and the use of partial coherent source in diffractive imaging experiments [15, 16, 17, 18].

 To accommodate these additional degrees of freedom during specimen recovery, ptychography IPR algorithms exploit the high redundancy of the dataset that comes from the repeated expression of the specimen information in multiple diffraction patterns. Previous measures of this redundancy use the constraint ratio: a number counting metric that quantifies the sufficiency of the measurement constraints [19, 8]. In a recent work we showed that the number counting metric for ptychography (the oversampling ratio) is independent of the illumination size and shape [12]. This work also showed that when a structured probe is employed in ptychography, the internal distribution of redundancy of the dataset cannot be characterise by a single number.

 In this paper we present a complementary view of the dataset redundancy, which is independent of the oversampling ratio and is capable of describing how the illumination structure affects the redundant expression of the specimen information. To accomplish this, we describe the experimental setup (comprising the illumination profile and sampling configuration) using a vector space model of ptychography since this allows us to readily discern their individual role on the dataset redundancy. It will become clear as we progress through this paper that the vector space quantity that encodes the redundant expression of the specimen information in the dataset is the Wigner distribution function (WDF) of the illumination. This new interpretation of the WDF puts it at the centre of data expression in ptychography, which strongly suggests that it should not only show up in the closed form solution of the method [20, 21], but also in the solutions provided by iterative algorithms [24]. It is however not clear exactly how the WDF features in current ptychography IPR algorithms. Thus, we present an iterative formulation of ptychography using the WDF in order to elucidate its role in IPR algorithms, especially the PIE algorithm and its variants.

## 2. Vector space model of ptychography

The experimental setup of ptychography is determined by the form of the illumination at the specimen plane and the sampling configuration (detector pixel and probe position distributions). The illumination profile comprises several aspects such as boundary conditions in the case of the top hat aperture setup, or magnitude and phase variations in the case of structured illumination. In order to account for the influence of all of these characteristics on the expression of the specimen information in ptychography, it is useful to write the 2D ptychography equation in a vector space form where the illumination profile and the sampling configuration parameterize the measurement bases for expanding the specimen. Although, we can only measure intensity values, it is useful to understand the interference across the four-dimensional complex-valued dataset in ptychography because the iterative phase retrieval (IPR) approach is by default a complex valued framework. Moreover, we will revisit the issue of the intensity measurements later on when the role of the constraint set become paramount. The value of the complex wave field impinging on the detector is given by the 2D ptychography equation:

$$M_{j,k} = \int d\boldsymbol{r} \exp[2\pi i \boldsymbol{u}_j \cdot \boldsymbol{r}] a(\boldsymbol{r} - \boldsymbol{R}_k) q(\boldsymbol{r}). \tag{1}$$

Here $j$ and $k$ are integer parameters that enumerate the measurements of the dataset, $\boldsymbol{u}_j$ is the detector pixel coordinate, $\boldsymbol{R}_k$ is the probe position vector, $a(\boldsymbol{r})$ and $q(\boldsymbol{r})$ are the real space representations of the illumination and specimen functions respectively. These functions lie on the specimen plane, which is spanned by the 2D real space coordinate $\boldsymbol{r}$. The values of $\boldsymbol{u}_j$ ($\boldsymbol{R}_k$) equal integer multiples of the average reciprocal (real) space sampling interval, $U$ ($R$). For the purpose of presenting the vector space framework of ptychography, Eq. (1) is the representation of the specimen function in a coordinate system where the measurement bases are given by $A_{j,k}(\boldsymbol{r}) \triangleq \exp[2\pi i \boldsymbol{u}_j \cdot \boldsymbol{r}] a(\boldsymbol{r} - \boldsymbol{R}_k)$ and the projection coefficients are $M_{j,k}$. Since we are dealing with complex quantities, it is useful to adopt the Dirac notation as this greatly simplifies the analyses. The 2D ptychography equation in the Dirac notation is

$$M_{j,k} = \langle A_{j,k}|q\rangle = \int d\mathbf{r}\, \langle A_{j,k}|\mathbf{r}\rangle\langle \mathbf{r}|q\rangle, \qquad (2)$$

where the real space representation of the measurement basis is defined as, $\langle A_{j,k}|\mathbf{r}\rangle \triangleq A_{j,k}(\mathbf{r})$, and the specimen transmission function is given by $\langle \mathbf{r}|q\rangle \triangleq q(\mathbf{r})$. We use the completeness relation, $\mathbf{1} = \int d\mathbf{r}\, |\mathbf{r}\rangle\langle \mathbf{r}|$, to derive the real space representation of the projection in Eq. (1).

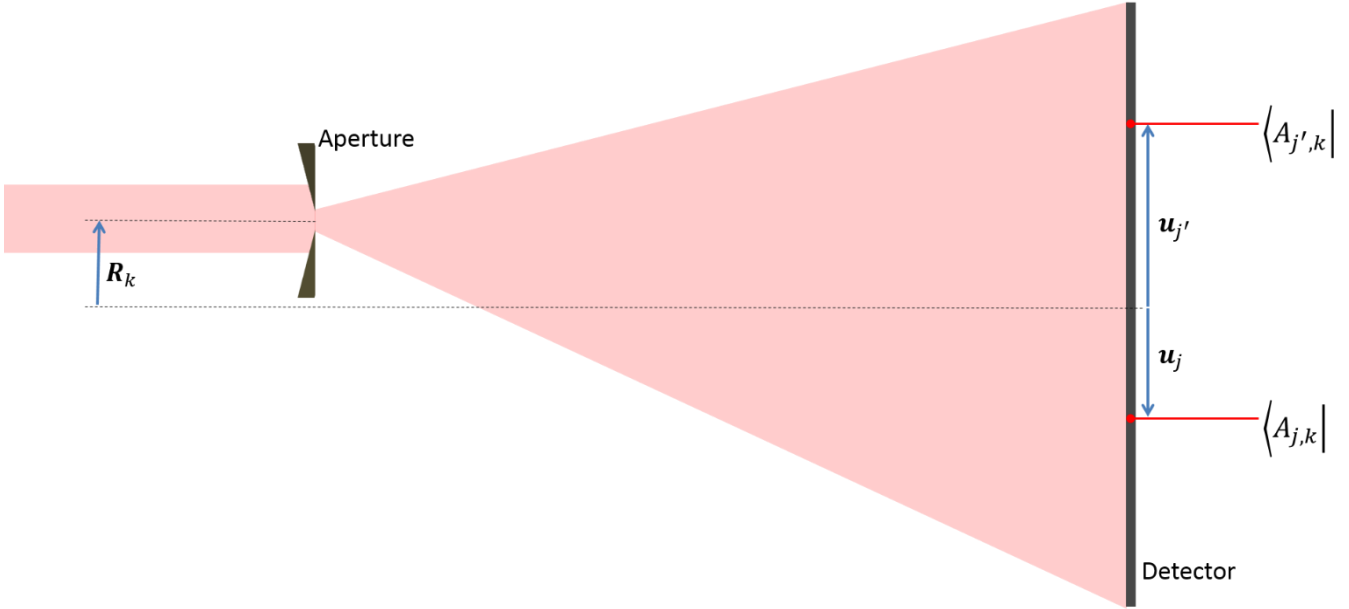

Figure 1 – (Color online) The vector space model of the experimental setup in ptychography. In this simple setup, the detector resides the Fraunhofer regime and the aperture is coincident with the specimen plane so that the illumination profile is the top hat aperture.

Figure 1 is a pictorial illustration of the correspondence between the measurement bases and the experimental setup in ptychography. The measurement basis $\langle A_{j,k}|$ encodes the relevant piece of information that photons (or electrons) that make up the illuminating beam arrive at the detector coordinate, $\mathbf{u}_j$, after traversing a localized real space region (with centre of coordinate $\mathbf{R}_k$) during the ptychography experiment. In this framework, the value of the intensity measurement at the coordinate $(\mathbf{u}_j, \mathbf{R}_k)$ is the expectation value of the specimen potential (represented by the operator $\hat{q} \triangleq |q\rangle\langle q|$) in the corresponding measurement basis, i.e.

$$I_{j,k} = \langle \hat{q} \rangle_{j,k} = \langle A_{j,k}|q\rangle\langle q|A_{j,k}\rangle$$

This is equivalent to the probability distribution of the probing radiation over the detector plane for the setup in Fig. 1 as long as the measurement bases are normalized, that is $\langle A_{j,k}|A_{j,k}\rangle = 1$ for all $j$ and $k$.

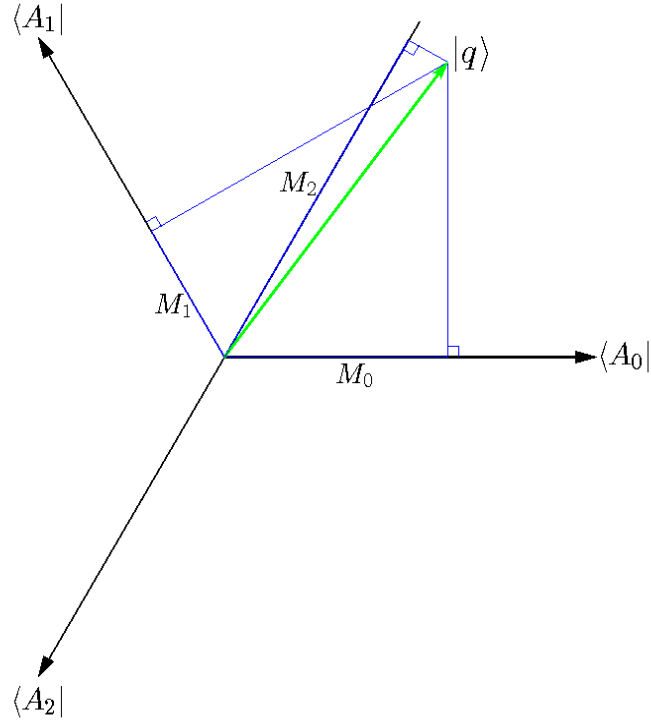

Figure 2 – (Color online) Illustration of the projection operation in Eq. (2) for a two-pixel specimen ptychography experiment. Here, the specimen $\langle r|q\rangle$ is represented by the column vector $[q_x \quad q_y]^T$ and the measurement bases are row vectors derived from the equation: $\langle A_j|r\rangle = [\cos(\phi_j) \quad \sin(\phi_j)]$, where $\phi_j = 2\pi j/3$ and $j$ takes on integer values in the range [0 3). Therefore, we can write the projection equation of Eq. (2) in matrix form as $M = Aq$ or less compactly as $M_j = \sum_r A_j^{\ r} q_r$, where the rows of the matrix are occupied by the measurement bases and $M_j$ equal the measurement values.

Let us consider a simple experiment illustrated in Fig. 2, where we collect three measurements ($M_j$) from a two-pixel specimen. The measurement value is the length of the projected component of the specimen along the measurement bases directions. An inspection of these directions (see Fig. 2) shows that when we employ more than two measurement bases (oversampling), the resulting data can be made to capture redundant information about the specimen provided we could identify at least two non-orthogonal pair. This simple setup captures a very important property about the real ptychography experimental setup, namely, the inverse operation to get back to the specimen can be carried out using scaled versions of the measurement bases. In other words, the dual or recovery bases are related to the measurement bases by a single scaling factor. Consequently, there is no need to perform a matrix inversion to get back to the specimen, as we will demonstrate shortly.

To recover the specimen from the measurements, we use the vector space expansion formula, $\boldsymbol{V} = \sum_i V^i \boldsymbol{e}_i$, where the measurements $V^i$ ($= \boldsymbol{V} \cdot \boldsymbol{e}^i$) are used as expansion coefficients in the recovery bases $\boldsymbol{e}_i$ (a.k.a the dual bases). In the Dirac notation the recovery equation for ptychography is

$$|q\rangle = \sum_{j,k} M_{j,k} |A_{j,k}\rangle. \qquad (3)$$

The recovery process for the two-pixel specimen experiment is illustrated in Fig. 3. The distance traversed in a given dual basis (or ket) vector direction is quantified by the value of the corresponding measurement.

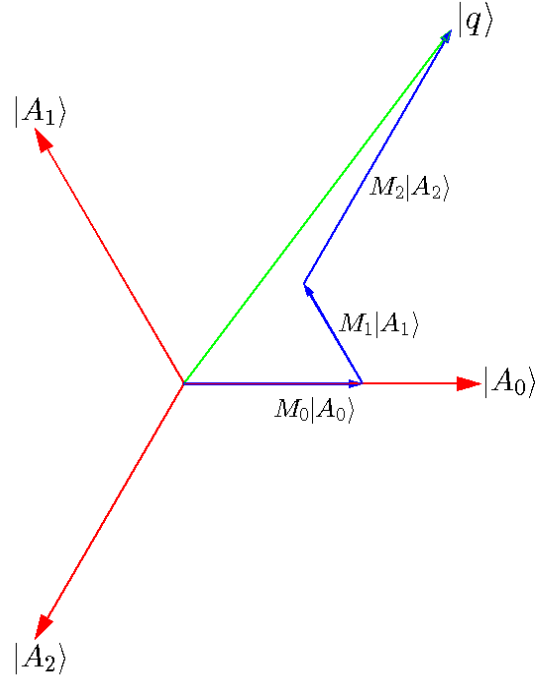

Figure 3 – (Color online) Illustration of specimen recovery (or deconvolution) in ptychography. We show here that deconvolution in ptychography is the weighted addition of the dual bases vectors, where the weighting factors are the measurement values and the entries of the dual basis vectors are proportional to the entries of the corresponding measurement basis vectors. In matrix notation, this operation is represented by matrix multiplication of the measurements ($M$) by the left inverse, $A_{\text{left}}^{-1}$. In ptychography, the left inverse is a scaled version of the Hermitian matrix $A^\dagger$ (or $A^T$ in the case of real-valued entries). The scale factor $\beta$ in this particular case is 2/3, but for the case of a general N-sided polygon, it is $2/N$. The column space of the left inverse matrix, $\beta A^T$, is spanned/occupied by the dual basis vectors—which we write using the Dirac notation as: $\langle r|A_j\rangle = \beta[\cos(\phi_j)\quad \sin(\phi_j)]^T$. In other words $q = \beta A^T M$, or equivalently $\mathbf{1} = \beta A^T A$. We know that the three measurement bases in Fig. 2 span the space of the two pixel specimen, since they are not collinear. To express this fact mathematically, we use the completeness relation, $\mathbf{1} = \beta A^T A = \sum_j |A_j\rangle\langle A_j|$.

To correctly recover the specimen, we require that the measurements bases span the space of the specimen. The sufficiency condition needed to satisfy this requirement is the completeness relation,

$$\mathbf{1} = \sum_{j,k} |A_{j,k}\rangle\langle A_{j,k}|,$$

where $U$ and $R$ equal to the sampling pitches for the reciprocal and real space windows that are connected by the Discrete Fourier Transform (DFT) equation. In a real experiment the size of the reciprocal space window is fixed by the angular span of the detector, whose sampling pitch is fine enough to ensure that the real space window contains the majority of the illuminating beam power. We confirm the validity of these constraints by performing the recovery (or deconvolution) operation in Eq. (3) using the completeness relation as follows:

$$|q\rangle = \mathbf{1}|q\rangle = \left(\sum_{j,k}|A_{j,k}\rangle\langle A_{j,k}|\right)|q\rangle = \sum_{j,k} M_{j,k}|A_{j,k}\rangle.$$

The real space representation of the specimen recovery in Eq. (3) is $\langle r|q\rangle = \sum_{j,k} M_{j,k}\langle r|A_{j,k}\rangle$, which after some rearrangement gives

$$q(\boldsymbol{r}) = \sum_k a^*(\boldsymbol{r} - \boldsymbol{R}_k)\psi_k(\boldsymbol{r}). \tag{4}$$

Here, the quantity $\psi(=\mathcal{F}^{-1}M)$ represents the exit wave in the specimen plane and its detailed form is given by the inverse DFT equation $\psi_k(\boldsymbol{r}) = \sum_j M_{j,k} \exp[-2\pi i \boldsymbol{u}_j \cdot \boldsymbol{r}]$, where $\mathcal{F}$ is the DFT operator. The inverse DFT kernel and the complex conjugate operation on the illumination ($a^*$) come from the relationship between the bra and ket representations in the Dirac notation, namely $\langle \boldsymbol{r}|A_{j,k}\rangle = \langle A_{j,k}|\boldsymbol{r}\rangle^*$. Eq. (4) provides the means for recovering the specimen in situations where the dataset is densely sampled (see the constraints on $U$ and $R$), the illumination is known *a priori* and the phase information is intact. The vector space approach to specimen recovery that led to Eq. (4) allows us to readily discern the striking fact that the recovery (or deconvolution) kernel is the conjugate of the projection (or convolution) kernel in ptychography.

Naturally, we need to understand and address issues such as the loss of phase information and sparse sampling, which are encountered in the experimental realization of ptychography. We will show later on that repeated application of Eq. (4) is sufficient to facilitate robust specimen recovery when the phase information is lost. As for the issue of sparse sampling of the probe position coordinate, previous methods employ a normalization scheme that divides the specimen function in Eq. (4) by the position dependent flux that traverses the specimen during the entire ptychography experiment [3, 22]. Here, we present a different approach that avoids the use of the division operation altogether. To do so, we view IPR in a broader context that deals with the partition of information about complex functions, where we recover the specimen information from an incomplete set of measurements. This incompleteness may present itself in a variety of ways such as: the loss of phase information (CDI and ptychography experiments); or the loss of the real (or imaginary) part of the function; or the loss of a subset of the complex wave field (sparse sampling); or a combination of some or all of these. When viewed in this way, we immediately realize that the central underlying theme of any IPR algorithm that can process different types of data (real, imaginary, phase and/or intensity measurements) is the requirement that the measurement bases are not all independent of one another (i.e. non-orthogonality). So that for example, the question of how phase information is encoded in a set of intensity measurements becomes a question about the linear dependence of the measurement bases corresponding to these intensity measurements. The vector space object that quantifies the amount of non-orthogonality is the inner product, which will be elaborated upon in the following section.

## 3. The WDF as a metric for comparing different experimental geometries

In the vector space model of ptychography, the amount of overlap between a pair of measurements that reside at say a fixed coordinate $(\boldsymbol{u}_0, \boldsymbol{R}_0)$ and a general coordinate $(\boldsymbol{u}_j, \boldsymbol{R}_k)$ is given by the inner product, $\langle A_{0,0}|A_{j,k}\rangle$, of the corresponding measurement bases. Using the real-space completeness relation, we write this inner product as

$$\chi_a(\mathbb{u}, \mathbb{R}) = \int dr\, a(r)a^*(r+\mathbb{R})\exp[-2\pi i \mathbb{u}\cdot r], \quad (5)$$

where $\mathbb{u} = u_j - u_0$, $\mathbb{R} = R_k - R_0$ and $u_0(R_0)$ is used as the centre of coordinate in reciprocal(real) space. Eq. (5) is the Wigner distribution function of the illumination (aWDF) that features prominently in the non-iterative solution of ptychography [20, 21]. We emphasize the difference of coordinate in its argument because the amount of overlap (or interference) cannot depend on the absolute value of the coordinate variables. In ptychography the only quantities that matter in this regard are the relative values of the coordinate, just like the case where only the relative phase values of a wave field is important. The aWDF measures the ability of the experimental setup to produce data that captures redundant information about the specimen, or more precisely, it is the redundancy distribution metric for the dataset.

For concreteness let us compare two different experimental setups using the aWDF. The first setup employs a curved illumination (whose reciprocal space aperture size is $K$) and the second setup employs a propagated aperture (made from the top hat aperture with a real space width, $D$). Without loss of generality, we use the one-dimensional version of these setups in the following discussions. The magnitudes of the aWDFs are shown in Fig. 4. We also include the aWDFs for the non-propagated version of these setups (the Airy disc and the top hat aperture) for reference.

The top- and bottom-right images of Fig. 4 show that the shearing process (from propagation) modifies the expression of information in the curved illumination setup a lot more than it does for the propagated aperture setup. This is because vertical shearing of the horizontal line ($\chi_a(\mathbb{u}, 0)$ from the Airy disc setup) gives a tilted line along which the measurements are highly correlated, see Fig. 4(f). Consequently, in the curved illumination setup the measurement basis for say the centre detector pixel points in the same direction as the measurement basis for a nearby detector pixel in an adjoining diffraction pattern of the dataset. The exact relationship is governed by the amount and sense (positive or negative) of the curvature of the illuminating beam. This effect is observed as translation of features in the detector plane as we scan through the probe positions, or equivalently, the motion of shadow images in the Gabor Hologram. This is however not the case for the propagated aperture setup where the minimally sheared vertical line $\chi_a(0, \mathbb{R})$ in Fig. 4(c) means that the strongest correlation occurs between measurements that reside in the same pixel location in the adjoining diffraction patterns. Consequently, the curved illumination setup is more sensitive to errors in the defocus parameter than the propagated aperture setup.

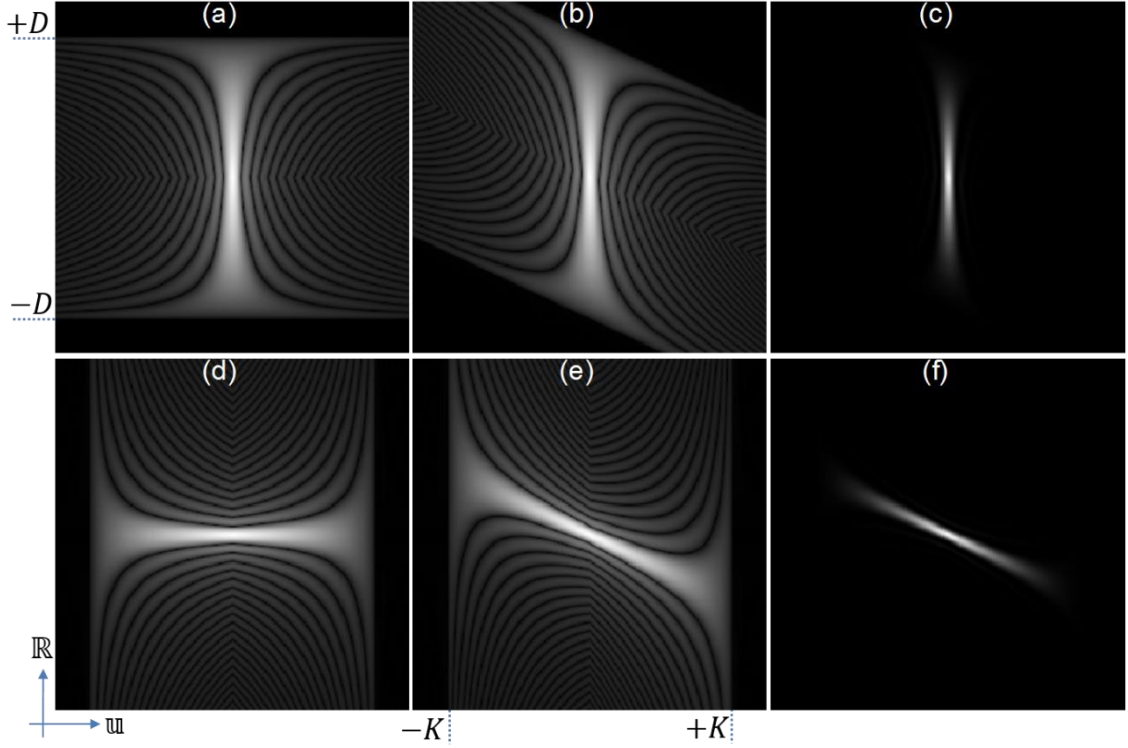

Figure 4 – Magnitude profiles of the aWDF for the one-dimensional version of the following setups: (a) Top hat aperture ($|\chi_a|^{1/2}$); (b) Propagated aperture ($|\chi_a|^{1/2}$); (c) Propagated aperture ($|\chi_a|^2$); (d) Airy disc ($|\chi_a|^{1/2}$); (e) Curved illumination ($|\chi_a|^{1/2}$); (f) Curved illumination ($|\chi_a|^2$). We display the intensity values in 4(c) and 4(f) to highlight the high magnitude region of the aWDFs. The horizontal and vertical directions correspond to the $\mathbb{u}$ and $\mathbb{R}$ coordinates respectively. In 4(a) the magnitude profile of the aWDF along the $\mathbb{u}$ coordinate direction is the scaled sinc function, $\text{sinc}(\rho\mathbb{u})$, where the scaling factor equals the triangular function of width $2D$, i.e. $\rho \triangleq \Lambda(\mathbb{R}/D)$. This setup has the kronecker delta aWDF, $\chi_a(\mathbb{u}, \mathbb{R}) = \delta(\mathbb{u}, \mathbb{R})$, when sampled at the Nyquist limit ($R = D$ and $U = 1/D$) because the real space sampling pitch picks out non-zero values of $\mathbb{R}$ that lies outside (above and below) the triangular envelope of the aWDF and the detector sampling pitch selects non-zero values of $\mathbb{u}$ that coincide with the zeros of the sinc function, $\chi_a(\mathbb{u}, \mathbf{0})$. Consequently, the corresponding intensity measurements capture no redundant information about an extended non-sparse specimen. An inspection of 4(a) to 4(f) shows that the aWDFs for the other setups are related to the top hat aWDF by: a vertical shearing operation (propagated aperture), a 90° rotation (Airy disc) and a 90° rotation followed by a vertical shear (curved illumination).

Furthermore, the range of the interference in real (or reciprocal) space is fixed by the sharp boundaries that demarcate the outer region of zeros from the interior structure of the aWDF. This boundary is defined by two vertical lines in the case of the Airy disc (and curved illumination) as shown in Fig. 4(d) (and 4(e)), which tells us that any measurement pair that are separated by a reciprocal space separation $K$ (an angular equivalent of twice the numerical aperture of the condenser lens) or more do not have the opportunity to interfere. For the top hat aperture setup, the interference condition is enforced by the real space boundary conditions of the aperture. Consequently, the sharp boundaries that separate the outer region of zeros from the interior structure of the top hat aWDF is defined by two horizontal lines. These lines are located at $\pm D$, where $D$ is the width of the top hat aperture. Intuitively, this means that information from two distinctly non-overlapping regions in real space cannot capture redundant information about the specimen.

## 4. Specimen recovery using the aWDF

During the IPR specimen recovery process, we compute the estimated values of the detector wave field at each iteration step and use the diffraction pattern to correct the magnitude of this wave field. In ptychography we do this for a set of diffraction patterns that are collected from a number of overlapping regions of the specimen. The rationale being that the magnitude constraint corrections from any one of these diffraction patterns should contribute toward the phase-profile update for adjacent probe-position wave fields. Since there is currently no transformation equation that connects the diffraction patterns of the dataset to one another, we use the consistency constraint of the overlapping region—a single specimen function common to all the exit waves—to guess the update equation for the detector plane wave field. Unfortunately, this approach results in a multitude of update equations that do not exploit the dataset redundancy in the correct way during specimen recovery. It also obscures the fact that the update equation for the detector plane wave field needs to take priority over other update equations (including those for the specimen and probe) because they are encoded in it and can be extracted when necessary. Thus, we need to find the mathematical relationship that links the underlying complex measurements of the dataset. We accomplish this by projecting the specimen function in Eq. (3) on to the measurement basis, $\langle A_{\mu,\nu}|$, to get the following equation

$$M_{\mu,\nu} = \sum_{j,k} \chi_a(\mathbb{u}, \mathbb{R}) M_{j,k}. \tag{6}$$

where $\mathbb{u} = \boldsymbol{u}_j - \boldsymbol{u}_\mu$, $\mathbb{R} = \boldsymbol{R}_k - \boldsymbol{R}_\nu$. From (6) we deduce that the equation needed to correctly transform the feedback error (due to magnitude constraint application) from one dataset coordinate to another is

$$\delta M_{\mu,\nu} = \sum_{j,k} \chi_a(\mathbb{u}, \mathbb{R}) \Delta M_{j,k}. \tag{7}$$

We prefix 'δ' to the quantity that we wish to calculate whereas the 'Δ' symbol precedes the quantity that contains the input data constraint. By making the following correspondences: $\delta M \rightarrow M_{n+1} - M_n$ and $\Delta M \rightarrow \widetilde{M}_n - M_n$, where $M_n$ and $\widetilde{M}_n$ are the $n$-th guesses of the detector plane wave field before and after the magnitude constraints application: we arrive at the vector space description of iterative specimen recovery in ptychography. That is

$$M_{\mu,\nu;n+1} = M_{\mu,\nu;n} + \sum_{j,k} \chi_a(\mathbb{u}, \mathbb{R}) \Delta M_{j,k;n}. \tag{8}$$

The preceding equation implies that updating the phase information of a given diffraction pattern (in a single parallel ptychography iteration step) requires that we take the weighted sum of the errors resulting from magnitude constraint corrections, where the weighting factor is the aWDF. Equation (8) is the detector plane update equation for the parallel-PIE algorithm [2]. It directly implies the required update equation for the specimen ket vector:

$$|q\rangle_{n+1} = |q\rangle_n + \left(\sum_{j,k} |A_{j,k}\rangle\langle A_{j,k}|\right)|\Delta q\rangle_n$$

because if we move all the terms to the left-hand side (LHS) of Eq. (8) so that the RHS equals zero, and use the fact that the measurement bases have unit magnitude, we immediately require the ket vector update equation to make the LHS equal to zero as well. The completeness relation of the measurement bases is not enforced in the ket vector equation because it does not apply in the presence of sparse sampling. We get the specimen plane update equation by projecting the $(n+1)$th guess of the specimen ket vector on to the real space bases $\{\langle r|\}$, i.e.

$$q_{n+1}(r) = q_n(r) + \sum_k a^*(r - R_k)\Delta\psi_{k;n}(r). \tag{9}$$

where $\Delta\psi (= \mathcal{F}^{-1}\Delta M)$ represents the difference between the updated and previous values of the exit wave fields in the specimen plane. Note that when the dataset is densely sampled, the first and the third terms on the RHS of Eq. (9) cancel out, so that we revert to the specimen recovery equation in Eq. (4). Thus in the dense sampling regime, repeated application of Eq. (4) is a good algorithm for specimen recovery when the phase information is lost.

The route to the specimen update equation in Eq. (9) shows that the PIE algorithm uses a single horizontal slice of the aWDF, $\chi_a(\mathbb{u}, R_1 - R_0)$, to update the phase profile of the diffraction pattern that resides in the dataset coordinate, $R_0$, using magnitude corrections from the previously processed diffraction pattern with probe position, $R_1$. As the iteration progresses, this process is repeated for a set of sparsely sampled probe positions, until these magnitude corrections (or feedback errors) are negligible. This error minimization property of the PIE algorithm is closely related to the steepest decent method [2].

## 5. Specimen and probe recovery

So far we have only considered the case where the illumination is moved over the specimen in order to generate the data. Since this is equivalent to moving the specimen in the opposite direction under the fixed illuminating beam, we can swap the roles of the specimen and the illumination in Eq. (9), so that the qWDF, $\chi_q(\mathbb{u}, -\mathbb{R})$, takes on the role of the aWDF, $\chi_a(\mathbb{u}, \mathbb{R})$. The resulting update equation for the illumination is

$$|a\rangle_{n+1} = |a\rangle_n + \left(\sum_{j,k} |Q_{j,k}\rangle\langle Q_{j,k}|\right)|\Delta a\rangle_n$$

Since the preceding equation does not account for the fact that the specimen also changes as the iteration progresses, it is prudent to revisit Eq. (7) to ensure that the magnitude constraint errors transform correctly under the simultaneous update of the specimen and the probe (i.e. accurate update equation for the detector-plane wave field). When the dataset is densely sampled, the measurement basis made out of the specimen also spans the space of the illumination, i.e.

$$\mathbf{1} = \sum_{j,k} |Q_{j,k}\rangle\langle Q_{j,k}|$$

This means that the transformation equation that connects all the measurements of the dataset in Eq. (6) is simultaneously satisfied by the qWDF and the aWDF. Therefore, a more general form of the detector plane error in Eq. (7) is

$$\delta M_{\mu,\nu} = \frac{1}{2}\sum_{j,k}\left(\chi_a(\mathbb{u},\mathbb{R}) + \chi_q(\mathbb{u},-\mathbb{R})\right)\Delta M_{j,k}. \tag{10}$$

Again, we consider terms preceded by `$\delta$' as relating to quantities that we wish to know and in principle know how to manipulate using basic calculus. For example, the real space representation of the LHS of Eq. (10), $\delta\psi(=\mathcal{F}^{-1}\delta M)$, gives the change in the exit wave resulting from changes in both the specimen and illumination functions. Using the multiplicative approximation model for the exit wave, $\psi = aq$, we know that this change is given by the equation $\delta\psi = a\delta q + q\delta a$. Thus, a detailed expansion of the LHS of Eq. (10) is

$$\delta M_{\mu,\nu}^{LHS} = \frac{1}{2}\mathcal{F}_\mu\bigl(a(\mathbf{r} - \mathbf{R}_\nu)\delta q(\mathbf{r}) + q(\mathbf{r} + \mathbf{R}_\nu)\delta a(\mathbf{r})\bigr),$$

where we split the exit wave into two parts: $a(\mathbf{r} - \mathbf{R}_\nu)q(\mathbf{r})/2$ in the first term and $q(\mathbf{r} + \mathbf{R}_\nu)a(\mathbf{r})/2$ in the second term. This equation gives the requisite placeholders ($\delta q$ and $\delta a$ for the specimen and the illumination respectively) that store the values of the changes induced when the input data is applied. Our next task is to connect both $\delta a$ and $\delta q$ to $\Delta M$. This is where the vector space approach comes into the picture. Using the discrete version of the real space completeness relation, $\mathbf{1} = \sum |\mathbf{r}\rangle\langle\mathbf{r}|$, the RHS of Eq. (10) evaluates to

$$\delta M_{\mu,\nu}^{RHS} = \frac{1}{2}\mathcal{F}_\mu\left(a(\mathbf{r} - \mathbf{R}_\nu)\sum_k a^*(\mathbf{r} - \mathbf{R}_k)\Delta\psi_k(\mathbf{r}) + q(\mathbf{r} + \mathbf{R}_\nu)\sum_k q^*(\mathbf{r} + \mathbf{R}_k)\Delta\psi_k(\mathbf{r})\right),$$

A term-by-term comparison of the expanded version of the LHS and the RHS of Eq. (10) shows that the changes in the specimen and illumination functions resulting from any change in the diffraction patterns (due to say magnitude constraint corrections) are given by

$$\delta q(\mathbf{r}) = \sum_k a^*(\mathbf{r} - \mathbf{R}_k)\Delta\psi_k(\mathbf{r}). \tag{11a}$$

$$\delta a(\mathbf{r}) = \sum_k q^*(\mathbf{r} + \mathbf{R}_k)\Delta\psi_k(\mathbf{r}). \tag{11b}$$

We incorporate Eq. (11a) and Eq. (11b) into an iterative algorithm by defining the following update equations:

$$\psi_{\nu;n+1}(\mathbf{r}) = \psi_{\nu;n}(\mathbf{r}) + \delta\psi_{\nu;n}(\mathbf{r}); \tag{12a}$$
$$q_{n+1}(\mathbf{r}) = q_n(\mathbf{r}) + \delta q_n(\mathbf{r}); \tag{12b}$$
$$a_{n+1}(\mathbf{r}) = a_n(\mathbf{r}) + \delta a_n(\mathbf{r}): \tag{12c}$$

for the exit wave, the specimen and the illumination respectively. The detailed form of the term that updates the exit wave is

$$\delta\psi_{v;n}(\mathbf{r}) = a_n(\mathbf{r} - \mathbf{R}_v)\delta q_n(\mathbf{r}) + q_n(\mathbf{r} + \mathbf{R}_v)\delta a_n(\mathbf{r}).$$

Using Eq. (11) the running estimates of the specimen and the illumination are updated by $\delta q_n = \beta_q \sum a_n^* \Delta\psi_n$ and $\delta a_n = \beta_a \sum q_n^* \Delta\psi_n$ respectively. The parameter $\beta_{a(q)}$ is used to control the convergence rate of the algorithm and typically takes on values in the range (0, 1]. The simultaneous update scheme presented in Eq. (12a) to Eq. (12c) constitutes what we call the wigner-PIE (or wPIE) algorithm.

The central core of the wPIE algorithm, which is captured by the vector space equation in Eq. (10), shows that the aWDF and the qWDF act as conduits that correctly move information about magnitude constraint corrections from one diffraction pattern to another during IPR calculations. Moreover, a close look at the update scheme reveals that the wPIE algorithm merges several common and complementary ideas from different methods including the nonlinear optimisation approach [2], the difference map method [3], the ePIE algorithm [4] and the WDDC method [20].

## 6.    Conclusion

This work demonstrates that the vector space model of ptychography provides a comprehensive framework for characterizing the role of the illumination in relation to dataset redundancy. In this framework, the intensity values of the dataset are the expectation values of the specimen in the measurement bases, which are themselves made out of the illumination profile and the sampling configuration of the experimental setup. Using the Dirac notation we demonstrated that in ptychography, the entries of the projection (or convolution) kernel are complex conjugates of the corresponding entries of the recovery (or deconvolution) kernel, and that this relationship is built into any technique (such as time-frequency transform methods) that samples the data using canonical coordinates.

Furthermore, the vector space model of ptychography provides a new interpretation for the WDF, which reveals that the structure of the aWDF maps out the redundancy distribution of the specimen information across the dataset. As a result, the aWDF is a unique qualitative (and quantitative) tool for comparing different experimental setups. We have also shown that the centralised role of the WDF on data expression makes it the ideal pivot for the formulation of iterative ptychography and exploited this fact in the development of the wPIE algorithm. This explicit use of the WDF in iterative ptychography bridges the gap between the two formerly disparate areas (non-iterative and iterative solutions) of the method. It also ensures that the wPIE algorithm optimally extracts information from the dataset.